# Discussions on the spatial exponential growth of electromagnetic quasinormal modes

T. Wu,[1] J. L. Jaramillo,[2] and P. Lalanne[1*]

[1] Laboratoire Photonique, Numérique et Nanosciences (LP2N), IOGS- Université de Bordeaux-CNRS, 33400 Talence cedex, France

[2] Institut de Mathématiques de Bourgogne (IMB), UMR 5584, CNRS, Université de Bourgogne, 21000 Dijon, France

* philippe.lalanne@institutoptique.fr

**Abstract.** The temporal response of open systems is marked by damped oscillations. These oscillations, often referred to as 'ringings', are the signature of the decay of quasinormal modes (QNMs). A major research objective across various fields is to represent the response of open systems using QNM expansions, akin to the treatment of normal modes in closed systems. In electromagnetism, it is widely acknowledged that QNM expansions provide a relevant representation of the modal physics within the interior of compact resonators in free space, where QNMs form a complete set of the resonator. However, challenges emerge in the exterior of the resonator, where QNM fields exhibit exponential divergence, rendering QNM expansions incomplete. The divergence poses delicate mathematical issues that often lead to misinterpretations on the physics side. Hereafter, we analyze foundational concepts such as cavity perturbation theory and dissipative coupling between resonators. By studying model problems, we show that the exponential growth is physical and meaningful for understanding the interaction between remote electromagnetic bodies. The analysis consistently reveals that the coupling coefficients between QNMs of two distant bodies strengthen as the separation distance increases, therein challenging the intuition that distant bodies behave independently. These insights that shed light on the origin and implications of the divergence hold significant implications for understanding the ability of contemporary electromagnetic QNM theories in offering a modal representation of the physics in the open space surrounding resonator bodies. Our critical examination of these theories reveals the existence of two distinct perspectives and elucidates the significance of their contrasting viewpoints.

## 1. Introduction

Many concepts in physics rely on the idea of normal modes of conservative systems, for example, energy eigenstates, molecular orbitals, transitions between states, excitation energies... In the normal-mode analysis of self-adjoint problems, the spectrum of the operator is discrete and the wavefunctions describing the response of the system must vanish outside a finite region in space. The response of the system, as well as the initial excitation, can be represented as a sum over the normal modes which form a complete set. And this representation is unique.

When energy dissipation emerges through processes like absorption or leakage, the system becomes nonconservative. The corresponding operator becomes non-self-adjoint, and its spectrum purely continuous. The spectrum also includes an infinite set of quasinormal modes [1] with isolated, complex frequencies, in a way closely related to true normal modes, at least for resonances with small damping.

QNMs spread in physics well beyond the electromagnetic context preferentially discussed below and have garnered widespread interest in recent years in relation with prominent topics in quantum mechanics [2-4], black-hole theory [5-9] or hydrodynamics and turbulence [10,11]. In electromagnetism, they are electromagnetic fields, $\tilde{\mathbf{E}}(\mathbf{r})$ and $\tilde{\mathbf{H}}(\mathbf{r})$ with an $\exp(-i\tilde{\omega}t)$ dependence, which satisfy the source-free Maxwell equations, $\nabla \times \tilde{\mathbf{E}} = i\tilde{\omega}\boldsymbol{\mu}(\mathbf{r},\tilde{\omega})\tilde{\mathbf{H}}$ and $\nabla \times \tilde{\mathbf{H}} = -i\tilde{\omega}\boldsymbol{\varepsilon}(\mathbf{r},\tilde{\omega})\tilde{\mathbf{E}}$, where $\boldsymbol{\varepsilon}(\mathbf{r},\omega)$ and $\boldsymbol{\mu}(\mathbf{r},\omega)$ are the frequency-dependent permittivity and permeability tensors of the system. They satisfy the outgoing wave condition that specifies that nothing is supposed to come in from spatial infinity towards the resonator, at $|\mathbf{r}| \to \infty$. QNMs play a unique role to understand the optical properties of resonators [12-**Erreur ! Source du renvoi introuvable.**], e.g. plasmonic resonators that promote light-matter interaction in deep subwavelength volumes.

Hereafter, we will often refer to the textbook case of compact resonators in free space. Theses resonators are characterized by a closed and finite volume of Euclidean space, where the permittivity differs from that of the surrounding infinite background. Examples include 1D slabs, 3D spheres or ellipsoids in uniform media. We will designate the interior of the resonator as the set of points within its volume, the resonator exterior as points lying outside it, and the boundary as the closed surface that delineates the resonator volume.

One key consequence of dissipation is the exponential growth (or divergence) of QNM fields within the open space outside the resonator. The origin of the divergence is readily understood by considering that the exponential temporal decay ($\text{Im}(\tilde{\omega}) < 0$) of QNM fields over time at a particular point is counterbalanced by an exponential field growth over space at a particular time. At far distances from resonators in the free space, QNMs take on the form of a spherical leaky wavefront, $|\mathbf{r}|^{-1}\exp[-i\tilde{\omega}(t-|\mathbf{r}|/c)]$, propagating outward with a phase velocity $c$.

Despite notable progresses in advanced mathematical theorems [24-25], regularization formalisms [22,26-29] to mitigate the divergence impact, normalization theories [12,14] to address complex structures currently investigated in nanophotonics, there is a persistent defiance towards accepting the divergence. The latter often appears as a mathematical abstraction disconnected from tangible reality for non- specialists and keeps on posing conceptual challenges to experts.

The issue arises from the QNM inability to form a complete set for infinite Euclidian spaces, unlike the normal modes of closed and finite spaces. Indeed, representing the scattered field (i.e., the resonator response to an applied field), which remains finite across all space, as a summation of wavefunctions that diverge at infinity ($|\mathbf{r}| \to \infty$), is likely impossible. Nonetheless, despite this difficulty, there is a need to quantify the extent to which certain physical processes excite QNMs. Thus, a quantitative measure for such excitation is necessary. This task, however, proves to be surprisingly challenging.

Theoretical investigations on electromagnetic QNMs frequently refer to the textbook case of compact resonators in free space, for which QNM expansions result in a convergent series inside the resonator [30,12-**Erreur ! Source du renvoi introuvable.**,14]. Despite the existence of several formulas for the expansion coefficients, a consequence of a mathematical properties of QNMs often referred to as 'overcompleteness' in the literature [31-32], there is general agreement that the modal physics is adequately captured in the interior of compact resonators.

The challenge intensifies when dealing with resonators exteriors, where the QNM set is incomplete [33]. This makes it difficult to providing a modal explanation of phenomena occurring in these regions, such as the interaction between a resonator and another distant body. Moreover, real-world systems add another layer of complexity; resonators like photonic-crystal cavities are typically non-compact and often rest on substrates rather than in free space. In such cases, the QNM set is incomplete everywhere across all space.

Besides the inherent hurdle posed by incompleteness, it is worth noting that divergence presents significant obstacles to our physical intuition and occasions some perplexity whether it "really corresponds to any physical reality", as noted long ago [34]. The divergence may also lead to misconceptions.

To illustrate this point, let us consider cavity perturbation theory, a cornerstone concept of various branches of physics [35-37]. First-order cavity perturbation theory (FOCPT) is an undisputed classical result for perturbers inside compact resonators [31]. The difficulty emerges when perturbers are located outside (Fig. 1): as the distance between the perturber and the cavity increases, the QNM field experienced by the perturber diverges – as does the backaction exerted by the perturber on the cavity modes. Thus, the coupling coefficients between a distant perturber and all the QNMs of a resonator should diverge as the separation distance increases, contrary to our natural expectation that the coupling would decrease. In the recent literature, this discordance has been described as "a fundamental problem with currently adopted formulas using QNM perturbation theory, when perturbations are added outside the resonator structure" [38], prompting the authors to introduce a new perturbation theory with 'regularized QNMs' that do not diverge. We will see that FOCPT remains valid regardless the perturber position, whether inside or outside the resonator, even at large distance where divergence becomes significant.

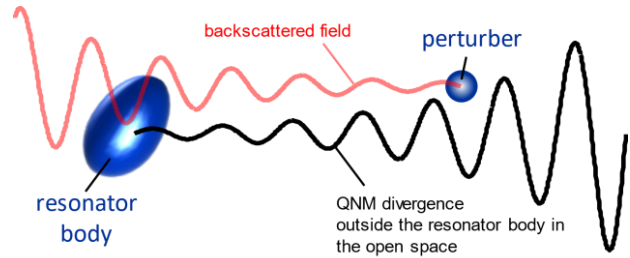

**FIG. 1.** First-order perturbation cavity theory (FOCPT) is an undisputed classical result of non-self-adjoint electromagnetism for perturbers placed inside resonator bodies [31]. However, a predicament emerges when considering perturbers located outside the resonator bodies: due to the divergence of the QNM field in the open space, the interaction between the resonators QNMs and the perturbers is exponentially growing up as the perturbers move away from the resonators.

In the following sections, we attempt to clarify the physical implications of the exponential divergence, in the temporal domain first and then in the frequency domain. Subsequently, we proceed to critique the abilities of present electromagnetic QNM theories in offering a modal representation of the physics in the open space surrounding the resonator. In particular, we reveal the existence of two distinct pictures and clarify the significance of their contrasted viewpoints.

The structure of the article is as follows.

In Section 2, we focus on the temporal response of resonators driven by a short pulse. We show that the divergence can be directly observed with simple experiments, highlighting the necessity of achieving a modal representation of physical phenomena in resonator exteriors, despite the incompleteness of QNM expansions there.

Following this brief excursion in the temporal domain, we shift our attention to the frequency domain in the next two sections.

In Section 3, we perform a comprehensive examination of FOCPT. We provide a mathematical and numerical evidence that FOCPT hold true for perturbations placed outside compact resonators, in the near field and even in the far field. We are led to the nonintuitive conclusion that QNMs are increasingly perturbed as perturbers move farther away.

Section 4 studies the dissipative coupling between two coupled resonators. It highlights that the divergence significantly affects the spectrum of the operator of the coupled system and that this influence increases with the separation distance between the resonators. This suggests that the QNMs of remote bodies, e.g. resonator or detectors, are entangled, challenging the intuition that remote bodies behave independently. With the help of Supp. Note 2, we then reconcile the nonintuitive result of Section 3 with our intuition that faraway perturbers exert vanishing influence.

These findings help us conducting a thorough evaluation of the predominant QNM theories [12-14,39-41] in Section 5 regarding their capability in capturing modal physics in open space. We identify a global weakness among all methods, especially in the most widely used method that fails to provide a modal description outside the resonator boundaries – a critical oversight, in our view. Conceptual challenges, lack of intuition, and theoretical deficiencies are unsurprisingly interconnected in this regard.

Finally, in Section 6, we offer an outlook of our conclusions.

The insights drawn from our conclusions within the realm of electromagnetism possess the potential to extend their influence into diverse fields. For instance, they could offer valuable assistance in appraising the underlaying physical character of "infrared" QNM instabilities encountered in black-hole theory. These instabilities which are conceivably triggered by changes in the faraway field asymptotics [6,42] or tiny "bumps" in the potential ("flea in the elephant" effect) [6,43-47] are crucial for interpreting present and forthcoming observations made by gravitational-wave interferometric antenna.

The following discussion will be supported by numerical illustrations conducted with simple settings for which large separation distances $d$ can be handled numerically: the double-barrier 1D problem that is a typical system discussed in many theoretical works on QNMs [1,6-8,12-**Erreur ! Source du renvoi introuvable.**] and a pair of 3D point-like dipolar polarizabilities. Nevertheless, we anticipate that all our illustrations well highlight the physics of complex environments currently investigated in nanophotonics.

## 2. The real and causal nature of the divergence

The physical reality of the diverging QNM field has occasioned some perplexity since more than a century. It was discussed as earlier as in 1900 by Lamb with the firm intention of showing that the exponential growth is observable [34,48]. Hereafter, we revisit Lamb's arguments to underscore the reality of divergence and emphasize the need to address it.

We consider the simple double-barrier problem for a normally incident Gaussian pulse with a central wavelength matched with the $m = 4$ resonance. The circles in Fig. 2a are computed with the temporal solver of COMSOL Multiphysics. This reference data shows the transmitted field $E(x_0, t)$ at a position $x_0$ chosen *outside* the resonator, $x_0 > L_1/2$. Initially null, the field first growths before exhibiting a quasi-monochromatic pattern that exponentially decays (additionally see the inset in long time settings).

The temporal response is easily interpreted by assuming a monomode response associated to the $m = 4$ resonance

$$E(x_0, t) \approx \text{Re}(A \exp[-i\widetilde{\omega}_4 (t - x_0/c)]) + E_{inc}(x_0, t), \tag{1}$$

the complex-valued excitation coefficient $A$ being fitted from the long-time settings. Except for an initial transient regime ($t < 75$ fs), the monomode response quantitatively reproduces the cavity ring-down of reference data. Note that Eq. (1) with the symbol '$\approx$' does not represent a single mode approximation of an exact expansion encompassing an infinite number of QNMs, as the latter expansion does not converge outside the resonator.

Naturally, we expect that, if the QNM field can be observed for a fixed position as a function of $t$, it may also be observed for a fixed time as a function of $x$. This hypothesis is verified in Fig. 2b, in which the slab response computed with COMSOL (circles) at $t_0 = 100$ fs is compared with the single QNM response of Eq. (1) (solid curve). We use the value of $A$ previously fitted. The spatial response carries clear evidence of the exponential growth of the QNM field, $\exp[i\widetilde{\omega}_4 x/c]$, therein highlighting that the divergence is not unphysical and *is even observable* [33], at least in a finite space interval for $x < 8$ µm.

Indeed, the temporal exponential decay (Fig. 2a) or spatial exponential growth (Fig. 2b) both come from one and only one physics, the excitation of a single QNM. Therefore, the exponential growth as $x$ *grows* is no more unphysical than that as $t$ *decreases*. Both divergences are not observed for all $t$ and $x$ owing to causality. In time, for a fixed $x_0$, no scattering happens before the pulse reaches the double barrier at time $t_{in}$. In space, for a fixed $t_0$, the exponential growth is observed only for $x < c(t_0 - t_{in})$, where the right-hand term represents the distance traveled by light between times $t_0$ and $t_{in}$. The spatial exponential growth is thus observable at *arbitrarily large $x$*, provided that the observation time is long enough (inset in Fig. 2b).

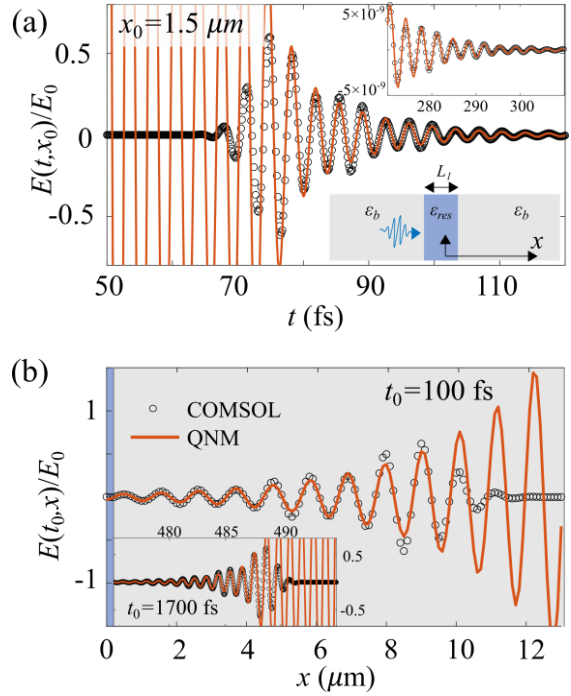

**FIG. 2.** The exponential divergence of QNM fields is observable under Gaussian pulse illuminations. **(a)** Temporal response $E(x_0, t)$ of a 1D barrier in air at a spatial coordinate $x_0 = 1.5$ μm outside the slab. The inset shows the temporal response at longer time. **(b)** Spatial response $E(x, t_0)$ of the barrier at time $t_0 = 100$ fs. The slab is shown with a blue rectangle. The inset shows the exponential growth at larger $x$'s for a longer time $t_0 = 1700$ fs. In **(a)** and **(b)**, full-wave data obtained with COMSOL are shown with circles and the monomode response of Eq. (6) is shown with a solid line. The pulse is defined by $E_0 \exp[ik_0(x - ct)] \exp[-(x - x_d - ct)^2/(2c^2\tau^2)]$, with $k_0 = 2\pi/1068$ nm$^{-1}$, $c\tau = k_0^{-1}$ and $x_d = -20$ μm. The permittivity and thickness of the barrier are $\varepsilon_{res} = 16\varepsilon_0$ and $L_1 = 400$ nm, and $\varepsilon_b = \varepsilon_0$.

## 3. First-order cavity perturbation theory for perturbations in the open space

Perturbation theory remains actively and heavily researched across multiple disciplines [8, 12,38,49-51]. It comprises methods for finding an approximate solution to a problem, by starting from the exact solution of a related, simpler problem. The solution is expressed as a power series in a small parameter, and the FOCPT corresponds to the first term of the series. It predicts the resonances of new (slightly perturbed) problems from the resonances of the initial (unperturbed) ones.

In electromagnetism, FOCPT provides a simple form of the resonance-frequency shift $\text{Re}(\Delta\widetilde{\omega})$ and the decay-rate change $-2\text{Im}(\Delta\widetilde{\omega})$ up to a term scaling as $|\Delta\varepsilon_{per}|^2$

$$\Delta\widetilde{\omega} = -\widetilde{\omega} \int_{V_{per}} \widetilde{\mathbf{E}} \cdot \Delta\varepsilon_{per}(\widetilde{\omega})\widetilde{\mathbf{E}} \, dV + O\left(|\Delta\varepsilon_{per}|^2\right), \qquad (2)$$

where $\Delta\varepsilon_{per}(\mathbf{r}, \omega)$ is a tiny permittivity change in a finite volume $V_{per}$, $\Delta\varepsilon_{per}(\mathbf{r}, \omega) \neq 0$ for $\mathbf{r} \in V_{per}$ only (Fig. 3a), and $\widetilde{\mathbf{E}}$ stands for the electric field of the unperturbed *normalized* QNM with complex frequency $\widetilde{\omega}$. Note that the unit of normalized QNM fields is not V.m$^{-1}$ but We assume a non-magnetic perturbation, but this assumption has no other effect than simplifying the treatment.

The QNM normalization issue has been solved about ten years ago, first for compact resonators in free space [52] and then for the general case of resonators in nonuniform background [22,53], see the comprehensive discussion in [14]. Since then, Eq. (2) has become an undisputed "classical" result for perturbations placed inside compact resonator bodies and has triggered significant success [18,**Erreur ! Source du renvoi introuvable.**,31-32,54-55]. It has also been numerically validated numerically for perturbations in the near-field of resonators [56], and recently, for delocalized perturbations spanning over the entire open space surrounding the resonators [57-59].

For perturbers positioned outside the resonator body, possibly at asymptotically large separation distances from the resonators, Eq. (2) predicts that $\Delta\tilde{\omega}$ diverges since $\tilde{\mathbf{E}}$ divergences. Despite its seemingly surprising nature [38], Eq. (2) holds true as we will see.

**Proof of Eq. (2) for faraway perturbers.** We consider a 3D resonator on a layered substrate for which QNM expansions are incomplete [14]. We will thus show that FOCPT validity has nothing to do with the completeness of the QNM expansion. The demonstration is based on the divergence theorem. It extends previous theoretical results [56] by offering more mathematical rigor and explicitly considering the case of faraway perturbers. The divergence theorem is applied to two source-free solutions of Maxwell equations, the unperturbed QNM and the perturbed QNM in the presence of the perturber.

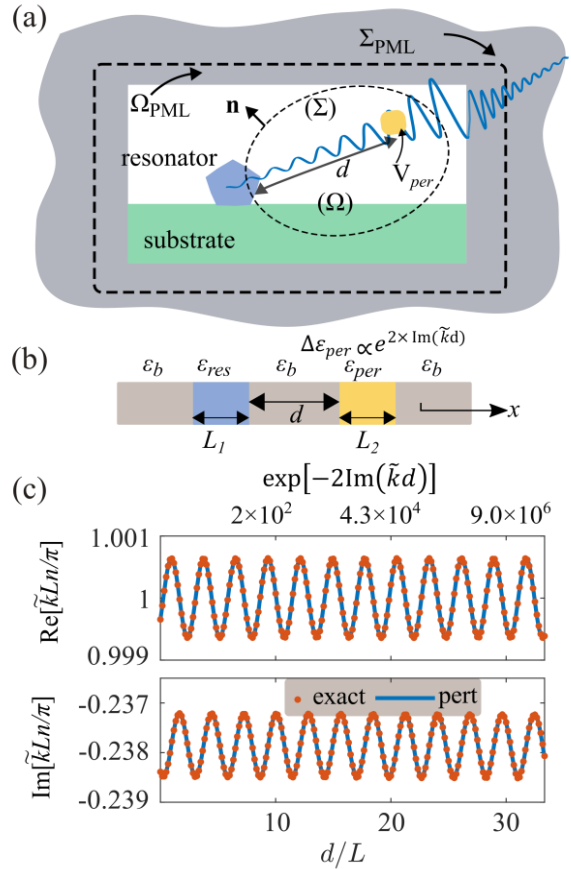

FIG. 3. FOCPT is valid for perturbations placed outside, even at arbitrarily large separation distance from the resonator. (a) For the mathematical proof, the (possibly nonuniform) open space around the resonator-perturbation system is surrounded by perfectly matched layers (PML) in which the divergent QNM field is damped to zero. (b) Sketch of the considered system for numerical tests: a 1D Fabry-Perot slab with refractive index $n = \sqrt{\varepsilon_{res}/\varepsilon_0}$ perturbed by another slab with refractive index $n_{per} = \sqrt{\varepsilon_{per}/\varepsilon_0}$. (c) Numerical test of Eq. (1) for large separation distances, $d$, and for the fundamental QNM with a single antinode of the QNM field ($\tilde{k}L_1 n =$

$\pi$ ). Other parameters are $L_1 = L_2 = L$ , $\varepsilon_{res} = 7.84\varepsilon_0$, $\varepsilon_b = \varepsilon_0$ , and $\varepsilon_{per} = \varepsilon_b + \Delta\varepsilon_{per0}e^{2\times\text{Im}(\tilde{k}d)}$ with $\Delta\varepsilon_{per0} = 4 \times 10^{-3}\varepsilon_0$ and $\varepsilon_0$ the vacuum permittivity. The solid blue curves are obtained with Eq. (1) and the red dots are numerical data obtained from a pole calculation of the scattering operator of the perturbed system.

To implement the divergence theorem, we first choose a separation distance $d$ between the perturber and the resonator (Fig. 3a). The distance may be arbitrarily large but is fixed. We then use the method of complex scaling [1,14,60] and surround the open space by a perfectly-matched layer (PML), i.e. a rotation of the coordinate in the complex plane that transforms the true exponentially-divergent original QNMs of the open space into regularized square-integrable QNMs, see Fig. 3a and Section 3.2 in Ref. [14] for a careful discussion of the electromagnetism case. The divergence theorem is then applied to two eigensolutions of the source-free Maxwell equations of the mapped space.

The first solution corresponds to a regularized QNM $(\tilde{\mathbf{E}}(\mathbf{r}), \tilde{\mathbf{H}}(\mathbf{r}))$ of the unperturbed resonator, $\nabla \times \tilde{\mathbf{E}} = -i\tilde{\omega}\boldsymbol{\mu}(\tilde{\omega})\tilde{\mathbf{H}}$, $\nabla \times \tilde{\mathbf{H}} = i\tilde{\omega}\boldsymbol{\varepsilon}(\tilde{\omega}) \tilde{\mathbf{E}}$, in which $\boldsymbol{\varepsilon}$ and $\boldsymbol{\mu}$ denote the possibly-dispersive permittivity and permeability tensors of the resonator on the substrate in the absence of perturbation and $\tilde{\omega}$ is the complex QNM eigenfrequency. The second solution, $\nabla \times \tilde{\mathbf{E}}' = -i\tilde{\omega}'\boldsymbol{\mu}(\tilde{\omega}')\tilde{\mathbf{H}}'$, $\nabla \times \tilde{\mathbf{H}}' = i\tilde{\omega}'[\boldsymbol{\varepsilon}(\mathbf{r}, \tilde{\omega}') + \Delta\boldsymbol{\varepsilon}_{per}(\mathbf{r}, \tilde{\omega}')] \tilde{\mathbf{E}}'$, corresponds to the perturbed QNM $(\tilde{\mathbf{E}}'(\mathbf{r}), \tilde{\mathbf{H}}'(\mathbf{r}))$ with an eigenfrequency $\tilde{\omega}'$, in the presence of the perturbation defined by the tiny permittivity change $\Delta\boldsymbol{\varepsilon}_{per}(\mathbf{r}, \omega) = \boldsymbol{\varepsilon}_{per}(\mathbf{r}, \omega) - \boldsymbol{\varepsilon}(\mathbf{r}, \omega)$, with $\Delta\boldsymbol{\varepsilon}_{per}(\mathbf{r}, \omega) \neq 0$ for $\mathbf{r} \in V_{per}$ only since the same PML is considered for the unperturbed and perturbed solutions.

We stress that the eigenfrequencies of the true and regularized QNMs are strictly equal and their respective fields coincide in the finite unmapped space (they only differ in the PML region $\Omega_{PML}$). Thus, since it is placed in the unmapped space $\Omega$, in the interior of the PML, the perturbation 'sees' the divergence of the true QNM.

We further consider a closed surface $\Sigma$ defining an inner volume $\Omega$ (Fig. 3a). For the moment, the surface is arbitrary; we just require that $V_{per} \subset \Omega$. We apply the divergence theorem to the vector $\tilde{\mathbf{E}}' \times \tilde{\mathbf{H}} - \tilde{\mathbf{E}} \times \tilde{\mathbf{H}}'$ to obtain

$$\oiint_\Sigma (\tilde{\mathbf{E}}' \times \tilde{\mathbf{H}} - \tilde{\mathbf{E}} \times \tilde{\mathbf{H}}') \cdot \mathbf{n} dS$$
$$= -i \int_\Omega \left[ \tilde{\mathbf{E}} \cdot \left( \tilde{\omega}\boldsymbol{\varepsilon}(\tilde{\omega}) - \tilde{\omega}'\boldsymbol{\varepsilon}(\tilde{\omega}') - \tilde{\omega}'\Delta\boldsymbol{\varepsilon}_{per}(\tilde{\omega}') \right) \tilde{\mathbf{E}}' - \tilde{\mathbf{H}} \cdot (\tilde{\omega}\boldsymbol{\mu}(\tilde{\omega}) - \tilde{\omega}'\boldsymbol{\mu}(\tilde{\omega}'))\tilde{\mathbf{H}}' \right] dV. \quad (3)$$

We now push the surface $\Sigma$ far inside the PML and consider the surface $\Sigma_{PML}$ (bold-dashed lines in Fig. 3a). Owing to the exponential decay of the regularized QNM fields in the PMLs, the surface integral on the left-side of Eq. (3) exponentially vanishes as $\Sigma_{PML}$ is pushed far inside the PML, and can be considered as null, letting us with

$$0 = \int_{\Omega_{PML}} \left[ \tilde{\mathbf{E}} \cdot \left( \tilde{\omega}\boldsymbol{\varepsilon}(\tilde{\omega}) - \tilde{\omega}'\boldsymbol{\varepsilon}(\tilde{\omega}') - \tilde{\omega}'\Delta\boldsymbol{\varepsilon}_{per}(\tilde{\omega}') \right) \tilde{\mathbf{E}}' - \tilde{\mathbf{H}} \cdot (\tilde{\omega}\boldsymbol{\mu}(\tilde{\omega}) - \tilde{\omega}'\boldsymbol{\mu}(\tilde{\omega}'))\tilde{\mathbf{H}}' \right] dV. \quad (4)$$

We denote the complex-valued resonance shift by $\Delta\tilde{\omega} = \tilde{\omega}' - \tilde{\omega}$. For FOCPT, we assume that a small perturbation $|\Delta\boldsymbol{\varepsilon}_{per}| \ll \mathbf{1}$ results in a frequency shift $\Delta\tilde{\omega} = O(|\Delta\boldsymbol{\varepsilon}_{per}|)$. This is the sole assumption of the proof. It is justified for QNMs that lie away from accumulation and exceptional points. Using Taylor series for the electric (or magnetic) field, $\tilde{\mathbf{E}}' = \tilde{\mathbf{E}} + O(|\Delta\boldsymbol{\varepsilon}_{per}|)$ and for the permittivity (or permeability), $\tilde{\omega}'\boldsymbol{\varepsilon}(\tilde{\omega}') = \tilde{\omega}\boldsymbol{\varepsilon}(\tilde{\omega}) + \frac{\partial[\omega\varepsilon(\omega)]}{\partial\omega}\Delta\tilde{\omega} + O(|\Delta\boldsymbol{\varepsilon}_{per}|^2)$, simple algebraic treatment from Eq. (4) leads to

$$\frac{\Delta\tilde{\omega}}{\tilde{\omega}} = -\frac{\int_{V_{per}} \tilde{\mathbf{E}} \cdot \Delta\boldsymbol{\varepsilon}_{per}(\tilde{\omega})\tilde{\mathbf{E}} \, dV}{\int_{\Omega_{PML}} \{\tilde{\mathbf{E}} \cdot \frac{\partial[\omega\varepsilon]}{\partial\omega}\tilde{\mathbf{E}} - \tilde{\mathbf{H}} \cdot \frac{\partial[\omega\mu]}{\partial\omega}\tilde{\mathbf{H}}\} dV} + O\left(|\Delta\boldsymbol{\varepsilon}_{per}|^2\right). \quad (5)$$

The denominator in the right-hand side of the equation represents the PML normalization factor [14,22]. For *normalized* QNMs, this factor is 1 and Eq. (5) reduces to Eq. (2).

The argument is very general, as it applies to arbitrarily large $d$'s and any QNM, not just the fundamental one. It highlights a very simple, albeit counterintuitive, result: the further the perturbation, the more it perturbs the resonator QNMs.

**Numerical test of Eq. (2).** This prediction is verified numerically for a 1D double-barrier slab perturbed by another slab (Fig. 3b). The normalized QNM field outside the resonator, $x > L_1/2$ (see the space coordinate in Fig. 3b), is a plane wave $\tilde{E}_y(x) = E_0 \exp(i\tilde{k}x)$, with $E_0$ fixed by the normalization and $\tilde{k} = \tilde{\omega}/c$. Equation (2) then reduces to $\Delta\tilde{\omega} = -\tilde{\omega} L_2 \Delta\varepsilon_{per} E_0^2 \exp[i\tilde{k}(2d + L_1 + L_2)] \text{sinc}(\tilde{k}L_2)$. We use exponentially-vanishing perturbations in the numerical test, $\Delta\varepsilon_{per} = 0.004 \times \exp(2 \operatorname{Im} \tilde{k}d)$, as $d$ increases, to remain in the range of validity of FOCPT regimes, which is a $O\left(|\Delta\boldsymbol{\varepsilon}_{per}|^2\right)$ with a prefactor that exponentially decays with $d$. On another side, we also compute $\Delta\tilde{\omega}$ directly by calculating the pole of the complex transmission coefficient of the perturbed 4-barrier system using $2 \times 2$ matrix product [61]. The comparison between the FOCPT prediction and full-wave numerical data shows a *perfect agreement*, even for $d$ as large as $32L$, for which $\exp(-2 \operatorname{Im} \tilde{k}d) \sim 10^7$ (Fig. 3c).

Let us note that opting for a 1D geometry to validate the formula is solely due to the possibility of conducting precise full-wave computations for faraway perturbers. The absence of peculiarities in 1D cases concerning the divergence leads us to anticipate that this test has broader implications, as demonstrated in the proof.

**Conclusion.** The mathematical derivation, obtained for the general case of arbitrary 3D systems, and the numerical test for a 1D system, firmly confirm that FOCPT is not restricted to the interior of compact resonators and remains valid for faraway perturbations. This unequivocally leads to the conclusion that *QNMs are increasingly affected as perturbers move farther away*. On another side, our everyday observation suggests that the optical responses of undisturbed resonators and those with distant perturbers appear identical.

This apparent dissonance is easily understood when we acknowledge that, in the numerical example, we intentionally ensure the perturber permittivity decreases exponentially as the distance $d$ increases, which keeps the system within the valid range of FOCPT. If the perturber permittivity had been held constant, as would be a natural approach, the exponential divergence of the QNM field would quickly invalidate the first-order perturbative regime, pushing the system into a strong coupling regime that requires higher-order (or even exact) perturbation theory. This regime is studied in detail in Suppl. Notes 1 and 2, as well as in the following Section. In summary, we demonstrate that there is no contradiction between the common assumption that distant objects do not interact and the fact that their QNMs can exhibit strong or even ultra-strong interactions.

## 4. Dissipative coupling between two faraway resonators

In this Section, instead of considering the coupling between a resonator and a perturber, we consider the corollary problem two coupled resonators and shows that the response of the coupled system is deeply influenced by the divergence. Dissipative coupling is a key concept in various fields, such as physics, electronics, and systems theory. It typically refers to a type of coupling between two bodies where some degree of energy is dissipated or lost in the process.

Understanding dissipative coupling is important in analyzing and designing systems where energy transfer and dissipation are key factors, as it can significantly impact system performance, stability, and efficiency. Our investigation highlights that the interaction between the resonators becomes increasingly pronounced as the separation distance grows, and the operator spectrum of the coupled system is characterized by numerous closely packed QNMs that are easily understood as Fabry-Perot QNMs.

**Operator spectrum of faraway resonant point-like dipoles.** We consider two identical (for simplicity) point-like dipolar oscillators with a Lorentzian resonant polarizability with resonant frequency $\widetilde{\omega}_0$: $\alpha(\omega) = \alpha_0/(\omega - \widetilde{\omega}_0)$. The value of $\alpha_0$ is chosen so that $\alpha$ corresponds to the polarizability of a 100nm-long gold plasmonic nanorod at a wavelength $2\pi c/\text{Re}(\widetilde{\omega}_0) = 665$ nm. The dipoles are both polarized parallel to the $z$ direction and are positioned on the $x$-axis with a separation distance $d$.

The induced dipoles are denoted $p_1$ and $p_2$. We further denote by $G$ the free-space Green-tensor component that connects the two dipoles. $G$ depends on the dipole separation distance $d$ and on the frequency $\omega$.

The QNMs of the coupled system with complex frequency $\omega$ are symmetric or antisymmetric states $(1, 1)$ or $(1, -1)$ with eigenfrequencies obtained by solving the source-free Maxwell equations

$$\begin{pmatrix} \omega - \widetilde{\omega}_0 & -\alpha_0 G(d, \omega) \\ -\alpha_0 G(d, \omega) & \omega - \widetilde{\omega}_0 \end{pmatrix} \begin{pmatrix} p_1 \\ p_2 \end{pmatrix} = 0, \tag{6}$$

where we can denote by $\kappa(\omega) = \alpha_0 G(d, \omega)$ the frequency-dependent coupling coefficient.

The spectrum of the coupled operator are the complex frequencies $\omega$ given by

$$\omega - \widetilde{\omega}_0 = \pm \kappa(\omega) = \pm \alpha_0 G(d, \omega). \tag{7}$$

In textbooks, Eq. (7) is usually solved in the static dipolar limit of small separation distances, $\widetilde{\omega}_0 d/c \to 0$, for which the evanescent near-field contribution of the Green tensor is predominant [62]. The Green tensor then becomes independent of the frequency and Eq. (7) becomes a quadratic equation in the variable $\omega$ with two complex roots $\widetilde{\omega}_b$ and $\widetilde{\omega}_{ab}$, corresponding to the bonding and antibonding states also known as subradiant and superradiant states.

Aside from the quasi-static limit, the complicated $\omega$-dependence of the 3D Green tensor, which includes an exponential term $\exp(i\omega d/c)$ for instance, should be considered. Then, Eq. (7) becomes a transcendental equation, and the number of solutions is not limited to two. It is rather infinite. The bottom panel in Fig. 4a shows the trajectories of all these roots in the complex plane as $d$ increases.

Let us first consider what is happening to the 'electrostatic' bonding and anti-bonding QNMs as $d$ increases. The top panel specifically depicts the dispersion of the bonding ($\widetilde{\omega}_b$) and antibonding ($\widetilde{\omega}_{ab}$) eigenfrequencies. The corresponding field maps are displayed in Fig. 4b for a few values of $d$. We can discern three distinct regimes. For small $d$'s, e.g. $d = 0.15\lambda_0$, the coupling is dominated by near-field interaction and the near-field interaction leads to a strong coupling with markedly different eigenfrequencies [62], $\widetilde{\omega}_b \neq \widetilde{\omega}_{ab} \neq \widetilde{\omega}_0$, see the blue marks in the upper panel of Fig. 4a. For intermediate $d$'s comparable to the wavelength, e.g. $d = 1.5\lambda_0$, the near-field interaction vanishes and the bonding and antibonding eigenfrequencies gently spiral around the resonator resonant frequency $\widetilde{\omega}_0$, a phenomenon known as the 'flea in the elephant' effect in black-hole theory [6,8,43-47]. The spiraling may result in transitioning from the strong coupling regime to the weak coupling regime, where $\widetilde{\omega}_b \approx \widetilde{\omega}_{ab} \approx \widetilde{\omega}_0$ (green marks in the upper panel of Fig. 4a). Finally, for large $d$'s, the coupling coefficient, $\kappa(\widetilde{\omega}_{ab})$ or $\kappa(\widetilde{\omega}_b)$, starts to diverge due to the presence of the exponential term, $\exp(i\omega r/c)$, in the Green function. As the result, the bonding and antibonding QNMs are propelled far away in the complex plane. They enter an 'ultra-strong' coupling regime. Note that the effect of the spatial divergence is also visible in the field maps in Fig. 4b for $d = 5.25\lambda_0$ and $7.5\lambda_0$.

In parallel to the bonding and antibonding migrations, the complex plane becomes increasingly populated with numerous Fabry-Perot quasinormal modes. Like the bonding and antibonding QNMs, all these modes rapidly migrate in the complex plane because their respective coupling coefficient diverges with $d$. Much like the 1D perturbed slab discussed in Suppl. Note 1, the abundance of these Fabry-Perot modes is directly proportional to the parameter $d$. These modes can be comprehended by considering that the dipolar resonators serve as resonant mirrors, causing only weak backward scattering (the mirror's reflectance decreases as $1/d$). The losses incurred during reflection must be compensated for by amplification ($\text{Im}(\omega) < 0$) to restore stationarity after one round trip. Because $d$ is relatively large, only a minor amount of amplification is required, and $\text{Im}(\omega) \propto \log(d)/d$, see Suppl. Note 2.

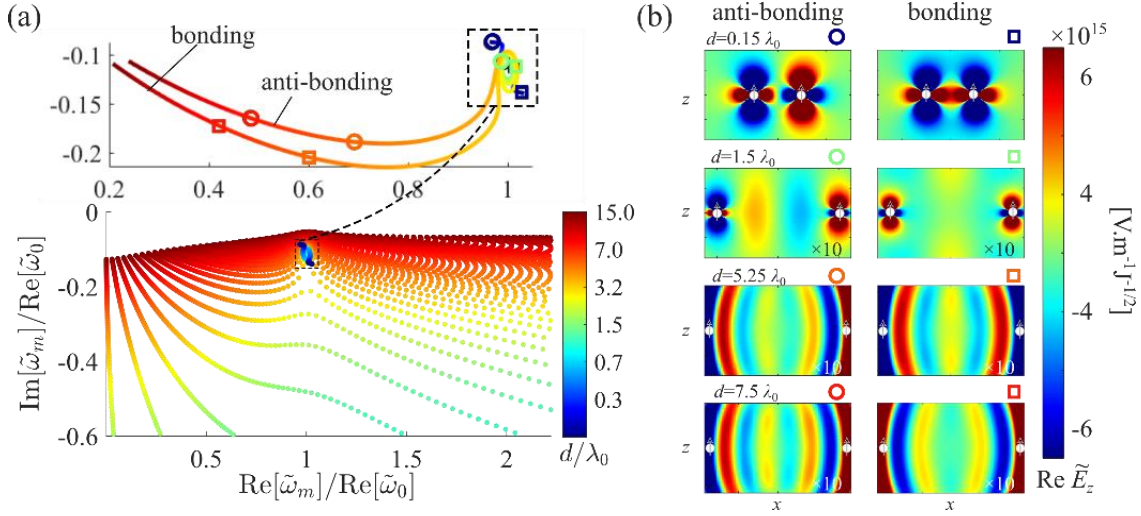

**FIG. 4.** Dissipative coupling between two point-like resonant dipoles. **(a)** Trajectories of the QNMs as a function of the separation distance $d$. The top inset is an enlarged view of the trajectories of the bonding and antibonding modes. The calculations are carried out for $\alpha_0 = (-1.5 + 0.15i) \times 10^{-17}$ s³·A²/kg and $\tilde{\omega}_0 = (2.8 - 0.31i) \times 10^{15}$ Hz. Electromagnetic interaction between two coupled point-like dipoles with resonant polarizabilities. **(b)** $z$-component of the normalized electric field, $Re(\tilde{E}_z)$, of the bonding and anti-bonding modes at various separation distances $d's$. The fields for $d = 1.5\lambda_0$, $5.25\lambda_0$, and $7.5\lambda_0$ cases are multiplied by a factor 10 for the sake of clarity. At $d = 0.15\lambda_0$, the coupling is dominated by near-field interaction. The latter becomes negligible for $d > 1.5\lambda_0$. For larger $d$'s, for instance $d = 5.25\lambda_0$ and $d = 7.5\lambda_0$, the interaction is dominated by $\exp(i\omega/cr)$ within the Green-tensor $G$. The white dots represent the dipoles.

**Response of the dipole pair under excitation by a monochromatic plane wave.** To better appreciate the role of the Fabry-Perot QNMs in the optical response of the coupled dipoles for increasing $d$ values, we reconstruct the dipoles response in the QNM basis. This theoretical work encompasses an original contribution on the normalization of QNMs of ensemble of Dirac resonant dipoles. The procedure, which may simplify the electromagnetic analysis of dipole ensembles coupled (or not) to resonators, e.g. atoms coupled to high-Q cavities [63] or molecules coupled to very small resonators [64], is reported in the Suppl. Note 3.

The results of the QNM reconstruction of the extinction of the dipole dimer are shown in Fig. 5. They are obtained for an illumination by a monochromatic plane wave $\mathbf{E}_{inc} = \mathbf{e}_z \exp[i\mathbf{k}_{inc} \cdot \mathbf{r}]$ with an incident angle of $\theta = 45°$ and for a plane-wave frequency $\omega$, approximately equal to the dipole resonance frequency, $\omega = \text{Re}(\tilde{\omega}_0)/0.98$. The reference results computed at real frequencies directly by solving the coupled system are shown with the orange dots.

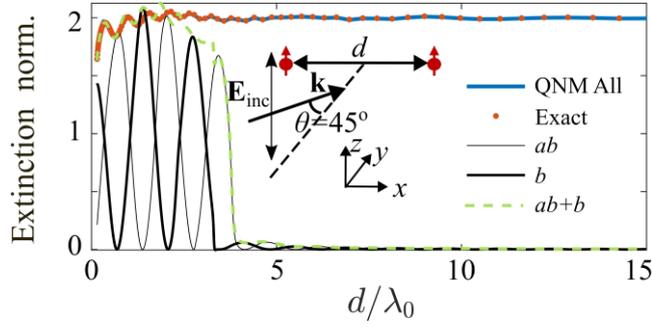

**FIG. 5.** Reconstruction of the normalized extinction of the dipole dimer as a function of the separation distance $d$ under monochromatic plane wave illumination. The QNM reconstruction (solid blue curve) obtained with all the QNMs within the spectral ranges $\text{Re}(\omega) \in [-29; 29]\, \text{Re}(\tilde{\omega}_0)$ and $\text{Im}(\omega) \in [-1.4; 0]\text{Re}(\tilde{\omega}_0)$ perfectly matches with the reference result (orange dotted curve) computed by directly solving the coupled system at real frequencies. The contributions of the bonding and anti-bonding QNMs are given by the thick and thin black curves, and their sum is shown with the thick green dashed curves. The bonding (b) and antibonding (ab) modes play a key role in the property on the coupled system for small $d$'s only. As $d$ increases, they strongly repeal each other because of the exponential divergence of their coupling strengths. Their contributions to the reconstruction vanish as $(\omega - \tilde{\omega}_b)^{-1}$ or $(\omega - \tilde{\omega}_{ab})^{-1}$. The calculations are carried out for $\alpha_0$ and $\tilde{\omega}_0$ provided in the caption of Fig. 4. The extinction is normalized by the extinction of a single isolated dipole.

The solid blue curve shows the extinction computed by retaining "all" (see the caption for details) the QNMs in the expansion. Since the dipoles are placed in a uniform background, the QNM expansion is complete (no branch cuts [14]) and as anticipated, a perfect agreement with the reference data is achieved. Note that this very good agreement validates the theory developed in Suppl. Note 3.

The figure also shows the contributions of the bonding and antibonding QNMs, with thick and thin black curves. For subwavelength $d$'s, the dimer response is shaped by the bonding and anti-bonding supermodes that result from the hybridization of the two oscillator modes: the sum (green dashed curve) of their respective contributions is nearly indistinguishable from the reference data for $d < 3.5\lambda_0$. This is well expected [62].

Interestingly, for large $d$'s, a strong or even ultra-strong coupling is implemented, and the bonding and anti-bonding modes are propelled far away from the dipole resonance frequency $\tilde{\omega}_0$ (upper panel in Fig. 4a). The propulsion phenomenon can be ascribed to the fact that, at $d > 3.5\lambda_0$, the divergent coupling between the two dipoles becomes significant. The propulsion is so extensive that the contributions of these two modes to the extinction become negligible for large $d$'s − (b) and (ab) curves in Fig. 5. Instead, the complex plane becomes densely populated by a myriad of Fabry-Perot modes, which predominantly influence the reconstruction of the field scattered by the oscillator dimer at large $d$'s (blue curve in Fig. 5).

**Discussion.** The earlier examinations of the dipole dimer and the perturbed resonator in Suppl. Note 1 align closely. This alignment is also consistent with the conclusion drawn in Section 3, indicating that QNMs are increasingly influenced as perturbers move farther away. The divergence shapes the complex-frequency spectra of perturbed (Fig. S2.1) or coupled (Fig. 4) operators. For large values of $d$, it eliminates the renowned influence of the bonding and antibonding modes on the response of resonant dimers and is responsible for the emergence of numerous Fabry-Perot QNMs, all with a large $Q$.

The complex-frequency spectra presented in Figs. 4 and S2.1 challenge our intuition, which suggests that the field scattered at real frequencies for 3D bodies decrease as $1/d$, implying that resonators become uncoupled as $d$ increases. However, distant perturbers indeed perturb, and distant resonators do not behave independently. If that were the case, the spectrum of the perturbed operator would solely consist of the unperturbed QNMs (depicted by the red stars in

Fig. S2.1), and the spectrum of the coupled operator (Fig. 4) would be governed by two identical eigenvalues at the resonant frequency $\widetilde{\omega}_0$ of the dipoles.

Reconstructions with numerous QNMs are somewhat impractically complicated in comparison to direct calculation. Nevertheless, without any intention of exaggerating our observations, the reconstructions presented in Figs. 5 and S2.3 unmistakably confirm the absence of any inconsistency with divergence. This aspect is expected to be relevant in more practical scenarios, where dissipative couplings occur between two or more bodies with subwavelength or wavelength-scale separations, a common occurrence in nanophotonics [65-66].

It is worth noting that, in the context of black hole theory, studies on long distant couplings at black hole horizons are predominant [43-47], which adds nuance to the issue of usefulness. With this regard, it would be interesting for further studies to consider the possible relation between the present Fabry-Perot modes and the Regge QNMs (not to be confused with Regge poles) in [1].

## 5. Management of divergence in QNM electromagnetic theories

So far, we have highlighted the widespread presence of QNM divergence in the physics of electromagnetic resonant systems, both in the temporal and complex-frequency domains. As previously noted, QNM expansions are incomplete in the open space outside the resonator, and this complicates the modal interpretation of phenomena involving electromagnetic fields therein.

In this Section, building upon the insights gleaned from the previous analysis, we closely examine the capabilities of prevalent theoretical and numerical methods in providing pertinent modal interpretations. We uncover significant limitations across all methods, arising from various causes, and highlight two primary representations, termed 'pointwise' and 'normal-like', which offer interesting, albeit quite distinct, viewpoints.

The examination is thorough, encompassing considerations such as the uniqueness, physical significance, and limitations of modal expansion, as well as both resonator interiors and exteriors, and the presence or absence of layered substrates. This lays the groundwork for presenting our conclusions in the following section.

**Real-frequency Green-tensor reconstruction method.** The most widely used method in electromagnetics originates from initial studies conducted for *compact resonators*, such as spheres or ellipsoids, *in free space* [30,14,12]. For these particular geometries, QNM expansions are complete [30,14] within the resonator volume $V_{res}$ and the scattered field can be accurately reconstructed there with expansions that exclusively consist of a discrete set of QNMs

$$\mathbf{E}_{sca}(\mathbf{r}_i, \omega) = \sum_m \alpha_m(\omega) \tilde{\mathbf{E}}_m(\mathbf{r}_i), \tag{8}$$

with $\mathbf{r}_i$ the "inside" coordinate, $\mathbf{r}_i \in V_{res}$. We emphasize that completeness is valid only for resonators in uniform backgrounds. In the presence of a substrate, which is very common in practical scenarios, branch cuts emerge in the complex plane [14]. These branch cuts necessitate additional computations, typically on a complexity scale similar to incorporating numerical modes in the PML regularization approach discussed below.

Owing to the uniqueness theorem [67], the internal field of Eq. (8) can be used to calculate the fields outside resonators. We symbolically denote the transformation that links the interior and exterior fields by $\mathbf{E}_{sca}(\mathbf{r}_o, \omega) = f_{i \to e}^{\omega}(\mathbf{E}_{sca}(\mathbf{r}_i, \omega))$, with $\mathbf{r}_o$ the "outside" coordinate. It is important to emphasize that the transformation $f_{i \to e}^{\omega}$ is frequency-dependent, necessitating a new computation for each distinct frequency $\omega$, whereas the interior field is known at all frequencies since the excitation coefficients are all known analytically [12].

In the presence of a layered substrate, there are several branch cuts with non-vanishing contributions and the expansion in Eq. (8) becomes incomplete. This is actually true for most quasinormal-mode systems in photonics.

Nothing is wrong with the widely adopted approach. However, by essence, it gets around the challenge posed by the divergence by ignoring any modal representation beyond the resonator boundary.

Incidentally, it is worth noting that the $\alpha_m$ coefficients in Eq. (8) lack uniqueness due to the overcomplete nature of the QNM basis [30,14,12]. This suggests that the most widely used method in electromagnetics does not offer a unique representation of the modal physics inside the resonator, even for uniform backgrounds. However, we note that the literature does not report any inconsistencies regarding the modal representation utilizing the most frequently used expressions for the $\alpha_m$ coefficients.

Recently, in order to confer modal significance to the field outside the resonator, it has been proposed to apply the interior-to-exterior field transformation at the level of every individual QNMs [68]. Equation (8) then becomes

$$\mathbf{E}_{sca}(\mathbf{r},\omega) = \sum_m \alpha_m(\omega) \begin{cases} \tilde{\mathbf{E}}_m(\mathbf{r}_i) \\ \mathbf{E}_m(\mathbf{r}_o,\omega) \end{cases}, \tag{9}$$

with $\mathbf{E}_m(\mathbf{r}_o,\omega) = f_{i\to e}^{\omega}\left(\tilde{\mathbf{E}}_m(\mathbf{r}_i)\right)$. The field on the right of the parenthesis, called 'regularized' QNM in [68], consists of a pair of solutions to Maxwell equations. The interior solution corresponds to the QNM field with complex frequency $\tilde{\omega}_m$, while the exterior solution involves a current source oscillating at real frequency $\omega$, generated outside by an interior field oscillating at $\tilde{\omega}_m$.

Equation (9) may allow faster computations when many computations at different frequencies are required. However, let us note an important issue when attempting to assign a significance to the nature of the 'regularized' QNMs [68]. Indeed, the later lacks integrity when assessed individually. This assertion can be readily comprehended by acknowledging that when a field oscillates at different frequencies outside and inside a compact body, the requisite interface conditions for tangent electromagnetic fields at the body boundary is not satisfied [14]. Only when considering the collective response of all "regularized" QNMs does integrity emerge. Thus, albeit quite natural, directly linking the modal physics outside resonators to the modal decomposition inside resonators relies more on arbitrariness than on a physically supported deduction. This will become evident with the discussion in Section 6.

**PML-regularized QNMs.** We now examine the PML-regularization approach [22,69-74,14]. Originating in quantum mechanics in the late 1970s, the approach was later extended to electromagnetism with the advent of PMLs [14]. It employs complex "scalings" to map the actual divergent QNMs of open space onto square-integrable regularized QNMs. In computational electromagnetism, the preferred mapping involves finite-thickness PMLs which surround an unmapped space ($\Omega$ in Fig. 3a) and are delimited by perfectly conducting metalic walls for instance.

The situation shares similarities with that encountered in the normal-mode theory of closed systems and the completeness of the scattered field expansion is guaranteed (in general, see Section 3.4.3 in [14]) in the whole computational domain

$$\mathbf{E}_{sca}(\mathbf{r},\omega) = \sum_m \alpha_m(\omega)\tilde{\mathbf{E}}_m(\mathbf{r}) + \sum_p \alpha_p(\omega)\tilde{\mathbf{E}}_p^{num}(\mathbf{r}), \tag{9}$$

provided that numerical eigenmodes $\tilde{\mathbf{E}}_p^{num}$ are included in the expansion. These modes are computed similarly to the QNMs as source-free solutions of Maxwell equations in the whole computational domain. The key distinction between numerical modes, $\tilde{\mathbf{E}}_p^{num}$, and QNMs, $\tilde{\mathbf{E}}_m$, lies in the fact that the numerical modes lack physical interpretation and depend on PMLs.

In the extended basis, accurate reconstructions have been obtained in either the spectral or temporal domains, even for intricate geometries such as resonators featuring dispersive materials positioned on layered substrates [14,69,74]. Notable success has also been achieved in investigating the coupling between resonators and perturbers [51,56-59] or among multiple resonators [75]. A unique force of the approach is its ability to offer an expansion that is complete, even when a layered substrate is present (the branch-cut contributions are accounted for with numerical modes). The expansion coefficients are also determined in a unique way, even for dispersive media if the expansion in Eq. (9) is generalized to quadri-QNMs formed by electric, magnetic and auxiliary fields, refer to Eq. (5) in [69].

We will designate this method as the "normal-like" approach since the scattered fields are expanded in a QNM basis with constant coefficients, within a closed and finite numerical space delimited by the metal walls.

Nonetheless, a constraint emerges in that attaining precision necessitates the incorporation of an increasing number of numerical eigenmodes as the separation distance from the resonator increases. As a result, convergence is practically limited to inner unmapped regions ($\Omega$ in Fig. 3) characterized by dimensions not larger than a few wavelengths.

Furthermore, the expansion lacks physical meaning when many numerical modes are involved. This issue becomes particularly significant in the time domain, which is readily accessible via Fourier transforms, as the expansion coefficients in Eq. (9) are known analytically for all frequency $\omega$. Although the temporal reconstruction remains exact even for complex geometries [69], it distorts the modal reality by overrating the divergence. Referring to Fig. 2b for illustration, the dominant QNM contribution typically resembles the red curves, whereas the actual QNM contribution should decay for large distances due to the finite velocity of wave-propagation in the open space. We note that the lack of physical meaning is also present in related approaches that do not explicitly involve numerical modes [73,76].

**Spatiotemporal settings**. The PML-regularization approach hinges on simple scalings that are solely contingent on spatial coordinates. In recent studies, spatiotemporal mappings involving the retarded time coordinate $u = t - r/c$ have been used to regularize the diverging field [7,77-80].

These mappings enforce causality as $t$ and $r \to \infty$. This implies that, in the temporal domain, each term of the temporal expansion, $\mathbf{E}_{sca}(\mathbf{r},t) = \sum_m \beta_m(t,\mathbf{r})\tilde{\mathbf{E}}_m(\mathbf{r})$, experiences an initial growth just beyond the resonator boundary due to the QNM divergence, followed by a decay as one moves farther away, eventually reaching negligible values at significantly large $\mathbf{r}$ distances – exactly as observed in the temporal response of Fig. 2. In other words, the finite velocity of wave-propagation is carefully handled by the retarded time coordinate.

The temporal expansion can be converted to the frequency domain using Fourier transforms [78]. Since this mathematical treatment preserves causality, it becomes viable to expand the field outside the resonator with QNMs [78,79]. However, this approach introduces a computational complication, as the expansion coefficients depend on the "outside" coordinate $\mathbf{r}_o$

$$\mathbf{E}_{sca}(\mathbf{r},\omega) = \sum_m \begin{cases} \alpha_m(\omega)\tilde{\mathbf{E}}_m(\mathbf{r}_i) \\ \beta_m(\omega,\mathbf{r}_o)\tilde{\mathbf{E}}_m(\mathbf{r}_o) \end{cases}. \tag{10}$$

Indeed, due to this dependence, the set of field distributions, denoted as $\mathbf{E}_m(\mathbf{r}_o,\omega)$, does not constitute a basis in the conventional sense, such as basis of functions exhibiting uniform convergence, as seen with normal modes of Hermitian systems, for instance. Rather, what we have is a 'pointwise' convergence, where the values of scattered fields for a specific $(\omega,\mathbf{r}_o)$ are effectively represented by a series of complex coefficients derived from the QNMs corresponding to that $(\omega, \boldsymbol{r}_o)$. Nonetheless, this 'pointwise' approach proves to be quite satisfactory.

The strict adherence to causality makes spatiotemporal settings a method of choice, and Eq. (10) has to be considered as a relevant expansion that is likely to adequately model the modal physics outside the resonator. However, the challenge lies in the fact that thus far, this approach has only been examined and validated for 1D systems or spherical scatterers (which inherently represent a 1D-like problem) [77,78]. Additionally, our experience with 1D system show that the expansion in Eq. (10) exhibits slow convergence for "outside" coordinates $\boldsymbol{r}_o$.

To sum up, not all approaches are equal regarding the divergence issue. However, they all encounter notable limitations. Conceptualization, theory, and numerical aspects are often intertwined, and it is unsurprising that the conceptualization challenges discussed in earlier sections manifest in both theoretical and numerical works.

## 6. Discussion

The intensity of waves scattered by local electromagnetic objects decreases with the distance from the objects and thus the electromagnetic interaction between distant objects diminishes as the separation distance between them increases. In this context, the exponential growth of QNM fields beyond the resonator's boundaries keep on casting doubt regarding the genuine implications, or even reality, of the growth [33], even in the recent specialized literature in electromagnetism [38].

The present discussion clarifies the issue by comprehensively integrating arguments supporting the physical reality of divergence. It shows that the divergence arises naturally as a causal consequence of the temporal exponential decay and can be observed with simple experiments [33,48] (Section 2). It also highlights that the divergence is deeply rooted in the physics of electromagnetic resonators, e.g. perturbation theory (Section 3) or dissipative coupling (Section 4). Our discussion also reconciles the apparent dissonance between the common assumption that distant objects do not interact and the fact that their QNMs exhibit strong interactions. It is worth noting that the FOCPT validity beyond the resonator boundaries justifies the use of complex-valued mode volumes [13] for interpreting experiments involving perturbers or emitters positioned in the near field. This resolves a previously unclear aspect in the literature [12-13,37-39,58] regarding the famous $Q/V$ factor.

We have encountered two distinct pictures regarding the modal representation of the physics within the open space surrounding the resonator.

The first picture relies on true QNM expansions with *constant* expansion coefficients. In this 'normal-mode-like' picture, emphasis is placed on the analogy to normal modes, viewing QNMs and their divergence as inherent properties of the underlying system, irrespective of the excitation field. Mathematical convergence is ensured by supplementing the QNM expansion with numerical modes, even in the presence of a substrate. However, when the contribution of numerical modes becomes dominant, particularly those associated with PMLs rather than branch cuts, the modal physics becomes inadequately conveyed, especially in the far field where QNM divergence prevails. Conversely, within the resonator and its near-field vicinity, the modal physics (e.g., resonator perturbations, Purcell effect with emitters near plasmonic antennas) appears consistently conveyed, without rupture. This perspective aligns with the observation that FOCPT holds everywhere in space (Section 3).

The other picture relies on a 'pointwise' expansion with coefficients that depend on the position of the observer. The excitation coefficient of every QNM at every frequency in the spectral domain is determined by the portion of the scattered signal capable of reaching the observer at the given time in the temporal domain (a property referred to as causality in [78]). This perspective likely offers the best portrayal of the modal physics for distant observers. However, it is noteworthy that spatiotemporal settings have only been validated for 1D systems or 3D spherical scatterers [5,7,77,78], and extending this to complex geometries like stratified substrates or non-spherical resonators could pose challenges. In such scenarios, establishing the retarded time coordinate may necessitate heightened implementation complexity [81].

Related discussions on this matter can be explored in the literature on black holes [5]. The question naturally arises: which perspective is 'correct'? We are inclined to argue that the distinction does not rest on being right or wrong, but rather on the viewpoint or approach to the problem, as long as the strengths and limitations of each perspective are recognized.

We hope that the present contribution, along with anticipated future enhancements in QNM theoretical and numerical methods, will help in understanding the consequences of divergence in the near, intermediate, and far-field zones.

## Acknowledgements

The authors acknowledge Benjamin Rousseaux, Mark Brongersma, Andrea Alù, Ling Miao and several other colleagues for helpful feedbacks during the preparation of the manuscript. PL acknowledges supports from the WHEEL (ANR-22CE24-0012-03) Project.